\documentclass[12pt]{article} 

\usepackage{latexsym} 	
\usepackage{amssymb}  	
\usepackage{amsbsy}
\usepackage{amsmath}
\usepackage{epsfig}     
\usepackage{graphics,color}
\usepackage{a4}
\usepackage{cite}

\newcommand{\eps}{\epsilon}
\newcommand{\vareps}{\varepsilon}

\newcommand{\bra}{\langle}
\newcommand{\ket}{\rangle}
\newcommand{\sgn}{\mbox{sgn}}
\newcommand{\half}{\frac{1}{2}}
\newcommand{\re}{\mbox{Re\,}}
\newcommand{\im}{\mbox{Im\,}}
\newcommand{\sign}{\mbox{sgn}}

\newcommand{\rmR}{{\rm R}}
\newcommand{\rmI}{{\rm I}}

\newcommand{\be}{\begin{equation}}
\newcommand{\ee}{\end{equation}}
\newcommand{\bse}{\begin{subequations}}
\newcommand{\ese}{\end{subequations}}
\newcommand{\bea}{\begin{eqnarray}}
\newcommand{\eea}{\end{eqnarray}}
\newcommand{\bean}{\begin{eqnarray*}}
\newcommand{\eean}{\end{eqnarray*}}
\newcommand{\nn}{\nonumber}
\newcommand{\hm}{\hspace*{-0.6cm}}

\newcommand{\bit}{\begin{itemize}}
\newcommand{\eit}{\end{itemize}}

\begin{document}

\title{ 
\vskip -100pt
{
\begin{normalsize}
\mbox{} \hfill \\
\mbox{} \hfill arXiv:1006.0332 [hep-lat]\\
\vskip  70pt
\end{normalsize}
}
\bf\large
Degenerate distributions in complex Langevin dynamics: one-dimensional
QCD at finite chemical potential
 }
                                                                                
\author{
 \addtocounter{footnote}{2}
 Gert Aarts$^a$\thanks{email: g.aarts@swan.ac.uk} 
 $\,$ and 
 K. Splittorff$^b$\thanks{email: split@nbi.dk}
 \\ \mbox{} \\
 {$^a$\em\normalsize Department of Physics, Swansea University,
 Swansea, United Kingdom}
 \\ 
 {$^b$\em\normalsize Niels Bohr Institute, Blegdamsvej 17, 
 DK-2100 Copenhagen \O, Denmark}
}
                                                                                
 \date{June 2, 2010}
                                                                                
 \maketitle
                                                                                
\begin{abstract}
 We demonstrate analytically that complex Langevin dynamics can solve the 
sign problem in one-dimensional QCD in the thermodynamic limit. In 
particular, it is shown that the contributions from the complex and highly 
oscillating spectral density of the Dirac operator to the chiral 
condensate are taken into account correctly. We find an infinite number of 
classical fixed points of the Langevin flow in the thermodynamic limit. 
The correct solution originates from a continuum of degenerate 
distributions in the complexified space.

\end{abstract}

\newpage
                                                                                
%\tableofcontents

% SECTION INTRODUCTION           

\section{Introduction}
\label{sec:Introduction}
\setcounter{equation}{0}

Random matrix theory and chiral perturbation theory have given deep 
insight into one of the most persisting problems of high energy physics, 
namely the {\sl sign problem}. The sign problem prohibits direct 
nonperturbative numerical simulations of strongly interacting matter with 
more quarks than antiquarks, since the measure on which the Monte Carlo 
method in lattice QCD seeks to sample the most important gauge field 
configurations is complex valued in the presence of a baryon chemical 
potential (for an excellent review see Ref.\ \cite{deForcrand:2010ys}).

Besides explaining why the quenched approximation fails at nonzero 
chemical potential \cite{Stephanov:1996ki}, random matrix theory and 
chiral perturbation theory have yielded highly nontrivial exact results 
for unquenched QCD at nonzero chemical potential $\mu$ 
\cite{Osborn:2004rf,Akemann:2004dr}.  From the viewpoint of the spectrum 
of the Dirac operator the effect of the chemical potential is always 
dramatic: The chemical potential breaks the antihermiticity of the Dirac 
operator. This drives the eigenvalues off the imaginary axis and makes the 
fermion determinant complex. In contrast, one of the few facts known about 
QCD at nonzero chemical potential is that the chemical potential has very 
little effect for low temperatures as long as $\mu$ is less than the scale 
set by the nucleon mass. This apparent mismatch between the strong effect 
of $\mu$ in the fermion determinant and the small effect of $\mu$ in the 
gauge averaged fermion determinant (the partition function) has been 
coined the Silver Blaze problem \cite{Cohen:2003kd}. The exact solutions 
for the unquenched average density of eigenvalues from random matrix 
theory and chiral perturbation theory show how this paradox is resolved 
\cite{Osborn:2005ss}: as the quark mass enters the band of eigenvalues in 
the complex plane, the average spectral density becomes complex valued and 
highly oscillatory. Since the period of the oscillations is inversely 
proportional to the space-time volume and the amplitude is exponentially 
large in the volume, the oscillations dramatically affect quantities such 
as the chiral condensate. In fact, the oscillations of the eigenvalue 
density are responsible for the discontinuity of the chiral condensate in 
the chiral limit, in strong contrast to the Banks-Casher relation valid 
for $\mu=0$ \cite{Banks:1979yr}.

The standard methods 
\cite{Fodor:2001au,Fodor:2002km,Allton:2002zi,Gavai:2003mf,deForcrand:2002ci, 
D'Elia:2002gd,D'Elia:2009qz,Kratochvila:2005mk,Alexandru:2005ix,Ejiri:2008xt, 
Fodor:2007vv} used to evade the sign problem in QCD (reweighting, Taylor 
series, analytic continuation, canonical ensemble, density of states) all 
become extremely hard to handle numerically when the sign problem is 
severe, i.e.\ when the cancellations due to the complexity of the weight 
change the partition function by a factor which is exponentially large in 
the volume. In order to understand the range of applicability, it has 
proven useful to study the interplay between the sign and Silver Blaze 
problems. One lesson that has emerged is the importance of the phase 
boundary of phase quenched QCD, i.e.\ the theory where the complex Dirac 
determinant is replaced by its absolute value, for simulations of full QCD 
at nonzero chemical potential \cite{Splittorff:2005wc}.

Complex Langevin dynamics \cite{Parisi:1984cs,Klauder:1983} differs from 
the approaches mentioned above in that importance sampling is not used. 
Instead the field space is complexified, which literally opens up new 
directions to evade the sign problem. Recently it has been shown that 
complex Langevin dynamics can solve the sign and Silver Blaze problems in 
the case of the relativistic Bose gas, i.e., a weakly coupled 
self-interacting complex scalar field at nonzero chemical potential, in 
four dimensions \cite{Aarts:2008wh,Aarts:2009hn}. Even though the sign 
problem is severe, the phase boundary of the corresponding phase quenched 
theory poses no obstacle. Promising results have also been obtained in 
heavy dense QCD and related models 
\cite{Aarts:2008rr}.\footnote{Early studies of complex Langevin dynamics 
can be found in, e.g., Refs.\ 
\cite{Klauder:1985b,Karsch:1985cb,Ambjorn:1985iw,Ambjorn:1986fz, 
Flower:1986hv,Ilgenfritz:1986cd}. Ref.\ \cite{Damgaard:1987rr} contains a 
further guide to the literature. Other recent work includes Refs.\ 
\cite{Berges:2005yt,Berges:2006xc,Berges:2007nr,Pehlevan:2007eq,Guralnik:2009pk}.
 }

 Complex Langevin dynamics is not without its problems, though. It has 
been known for a long time that instabilities and runaway solutions can 
result in a lack of convergence of the Langevin trajectories, which 
necessitates in some cases the use of an adaptive stepsize 
\cite{Ambjorn:1985iw,Ambjorn:1986fz}. Recently, it was shown how a 
straightforward implementation of an adaptive stepsize completely 
eliminates instabilities in heavy dense QCD and the three-dimensional XY 
model at nonzero chemical potential \cite{Aarts:2009dg}. Even when 
instabilities are under control, it is known that the dynamics can 
convergence to the wrong result \cite{Ambjorn:1986fz}. This problem has 
recently been studied in some detail in the case of simple models with 
complex noise \cite{Aarts:2009uq} and in the case of the three-dimensional 
XY model at nonzero chemical potential with real noise 
\cite{Aarts:2010aq}. Importantly, the conclusion reached in the latter 
work is that the erroneous convergence is independent of the strength of 
the sign problem.

In this paper we test the abilities of complex Langevin dynamics against 
the insights obtained about the Dirac spectrum and the sign problem: For 
QCD in one dimension we show that complex Langevin dynamics evaluates 
correctly the contributions from the extreme oscillations of the 
eigenvalue density of the Dirac operator. This provides further evidence 
that complex Langevin dynamics can solve the sign problem, and that the 
difficulties encountered in Refs.\ \cite{Aarts:2009uq,Aarts:2010aq} are 
not intrinsically related to sign problem but have a different origin. 
This clearly distinguishes this approach from the standard ones mentioned 
above.

One-dimensional QCD with $U(N_c)$ as gauge group contains many of the 
features which characterize the sign and Silver Blaze problems in 
four-dimensional QCD at low temperature \cite{Ravagli:2007rw}.\footnote{On 
the other hand, for $SU(N_c)$ the sign problem is not severe in 
one dimension \cite{Ravagli:2007rw}.} Moreover, it is exactly solvable, 
which makes it an excellent testground for new ideas 
\cite{Gibbs:1986xg,Bilic:1988rw,Lombardo:1999cz,Ravagli:2007rw,
Lombardo:2009aw}. In Ref.\ 
\cite{Ravagli:2007rw} it was shown that the spectrum of the Dirac operator 
in one-dimensional QCD at nonzero chemical potential is located on an 
ellipse in the complex plane. Since the chiral condensate can be viewed as 
the electric field originating from charges located at the positions of 
the eigenvalues, one would naively conclude that the chiral condensate is 
zero when the quark mass is inside the ellipse. However, the unquenched 
eigenvalue density on this ellipse is complex and rapidly oscillating: the 
correct chiral condensate, with a discontinuity when the quark mass goes 
through zero, emerges only when all highly oscillatory complex 
contributions are taken into account properly. Even though the Dirac 
spectrum in four-dimensional QCD is spread out into a band, it is exactly 
the same structure of the oscillations which is responsible for chiral 
symmetry breaking \cite{Osborn:2005ss}.

\begin{figure}[t]
\begin{center}
\epsfig{figure=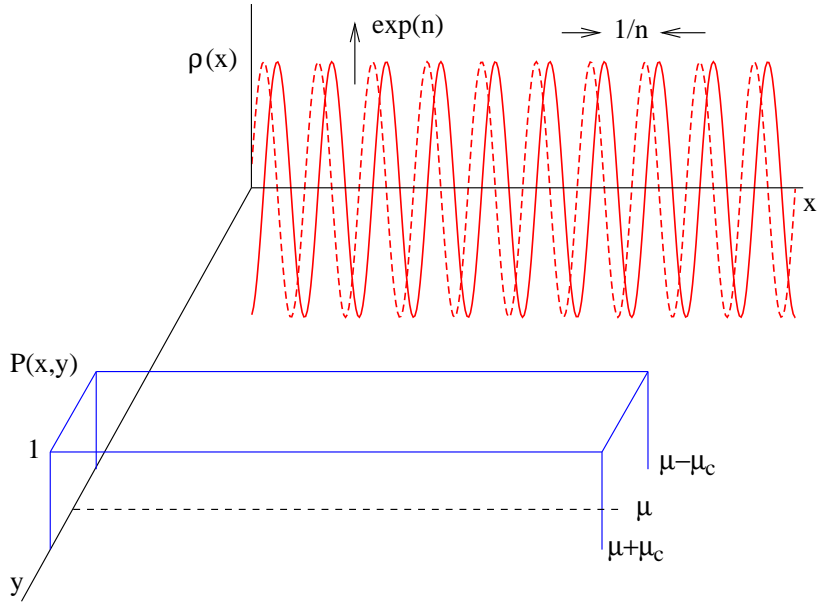,height=8cm} 
\end{center}
\caption{
 Sketch of how complex Langevin dynamics evades the sign problem in
 one-dimensional QCD, in the thermodynamic limit $n\to\infty$. The 
 oscillatory lines at $y=0$ represent the real (full) and imaginary   
(dashed) parts of the original complex weight $\rho(x)$. The uniform
 distribution $P(x,y)=1$ when $\mu-|\mu_c|<y<\mu+|\mu_c|$ and 0 elsewhere
 represents the distribution sampled by complex Langevin dynamics. All
 $y$ values in this region are equivalent, giving rise to the
 degeneracy mentioned in the title. The original sign problem is
 severe (mild) when $|\mu|>|\mu_c|$ ($|\mu|<|\mu_c|$).} 
\label{fig:dist}
\end{figure}

In this paper we establish a first link between the complex oscillations 
of the spectral density of the Dirac operator and complex Langevin 
dynamics, and demonstrate that the sign and Silver Blaze problems in one 
dimensional QCD are solved by complex Langevin dynamics.
We encounter a number of surprises not seen before. Most 
importantly, we find that complex Langevin trajectories depend on the 
initial conditions, even in the limit of infinite Langevin time: the 
dynamics is not ergodic. Nevertheless, all trajectories yield equivalent 
results, resulting in a degenerate set of stationary distributions in the 
complexified space. In the thermodynamic limit, this continuum of 
distributions becomes particularly simple and is sketched in Fig.\ 
\ref{fig:dist}. Moreover, it is known that classical fixed points can play 
an 
important role in localizing the dynamics in the complexified field space. 
In most models accessible to an analytical study, only a handful of fixed 
points exists. Instead, in the case studied here, we find an infinite 
number of stable and unstable classical fixed points in the thermodynamic 
limit.
 As in Ref.~\cite{Ambjorn:2002pz} where the factorization method was 
tested with random matrix results, our results demonstrate the usefulness 
of exact results as benchmarks for numerical methods which seek to deal 
with the sign problem.

The paper is organized as follows. After a brief review of the relevant 
analytic results in one-dimensional QCD in Sec.\ \ref{sec:1dqcd}, we set 
up and study the complex Langevin dynamics problem in Sec.\ \ref{sec:CL}. 
In Sec.\ \ref{sec:stat} we show that one stationary distribution of the 
corresponding Fokker-Planck equation can be found analytically for all $n$, 
but that, surprisingly, this distribution is not realized in the actual 
dynamics. This is explained in terms of classical flow and fixed points in 
Sec.~\ref{sec:flow}, where a larger set of solutions to the Fokker-Planck 
equation in the thermodynamic limit is given. It is demonstrated 
analytically that this continuum of degenerate distributions yields the 
correct result for the chiral condensate. Sec.~\ref{sec:concl} contains a 
brief summary and outlook.

\section{One-dimensional QCD and the sign problem}
\label{sec:1dqcd}
\setcounter{equation}{0}

We follow closely Refs.\ \cite{Bilic:1988rw,Ravagli:2007rw} and consider 
QCD with gauge group $U(N_c)$ in one dimension on a lattice with $n$ 
points. Here $n$ is assumed even throughout and is taken to infinity in 
the 
thermodynamic limit. The (staggered) fermions obey antiperiodic boundary 
conditions and chemical potential is introduced as usual 
\cite{Hasenfratz:1983ba}. We choose the gauge where all link variables are 
equal to unity, except at the final timeslice. The one flavour fermion 
determinant can then be written as
 \be
 \det[D(U)+m] = \det[e^{n \mu_c}+e^{-n 
\mu_c}+e^{n\mu}U+e^{-n\mu}U^\dagger],
\ee
 where $U$ is the remaining link variable and $\mu_c$ is related to the fermion mass $m$ via 
 \be
m=\sinh\mu_c.
\ee
 Since there is no Yang-Mills action in one dimension, the partition
function has the simple form 
\be
Z = \int_{U(N_c)} dU\, \det[D(U)+m] =
\frac{\sinh[n\mu_c(N_c+1)]}{\sinh(n\mu_c)}. 
\ee 
 One possible way to implement complex Langevin dynamics for this class of
theories is to evaluate the remaining group integral over the final link
variable using complex Langevin dynamics. Such an approach has been
explored in Ref.\ \cite{Aarts:2008rr} in the case of SU(3) and excellent
agreement with exact results has been obtained.
 Here we wish to make a connection between Langevin dynamics on one hand
and properties of the Dirac spectrum on the other hand. For that reason we
take a different route and first cast the chiral condensate as an integral
over the eigenvalue density of the Dirac operator. The resulting integral
is subsequently solved with complex Langevin dynamics.

 The partition function is independent of the chemical
potential. On the contrary, the eigenvalues of the Dirac operator $D(U)$, 
\be
\lambda_{k,l} = \half\left(e^{\frac{2\pi i(k+1/2) + i\theta_l}n+\mu}
-e^{-\frac{2\pi i(k+1/2) + i\theta_l}n-\mu}\right),
\label{lambdak}
\ee
depend on the chemical potential. Here $k=1,\ldots, n$, and 
$\exp(i\theta_l)$ with $l=1,\ldots, N_c$, are the eigenvalues of $U$.
The eigenvalues lie on an ellipse in the complex plane,
\be
\left(\frac{\re\lambda_{k,l}}{\sinh(\mu)}\right)^2
+
\left(\frac{\im\lambda_{k,l}}{\cosh(\mu)}\right)^2 =1.
\ee
Consequently, the eigenvalue density, 
\be
\rho(z;\mu)= \frac{1}{Z}\int_{U(N_c)} dU \, \det [D(U)+m] \, \sum_{k,l} 
\delta^2(z- \lambda_{k,l}),
\ee
depends on $\mu$. The chiral condensate (normalized with the 
one-dimensional volume), 
\bea
\nn
\Sigma = &&\hm \frac{1}{n} \frac{\partial}{\partial m} \log Z \\
= &&\hm \frac{1}{\cosh(\mu_c)} 
\left[ \left(1+N_c\right)\coth\left[\left(1+N_c\right)n\mu_c\right]
-\coth\left(n\mu_c\right)\right],
\label{eq:27}
\eea
is, however, independent of $\mu$ since the partition function
is. Expressing the chiral condensate as an integral over the
$\mu$-dependent eigenvalue density, 
\be
\Sigma = \int d^2z\, \frac{\rho(z;\mu)}{z+m},
\ee
 the $\mu$-independence of $\Sigma$ is far from obvious. We note that the 
condensate $\Sigma$ can be viewed as the electric field created by the 
charge density $\rho(z;\mu)$. Naively one would therefore expect the 
condensate to be zero when the mass $m$ is inside the ellipse in the 
complex plane, i.e.\ when $|m|<|\sinh(\mu)|$ or equivalently 
$|\mu_c|<|\mu|$. However, this is in contradiction with the known 
$\mu$-independence of the condensate.
 This is the {\sl Silver Blaze problem} in one-dimensional QCD: it 
illustrates the severe problems encountered when lattice QCD is applied 
for nonzero values of the chemical potential. It is thus of practical 
interest to understand if complex Langevin dynamics is able to reproduce 
this $\mu$-independence.

\begin{figure}[t]
\begin{center}
\epsfig{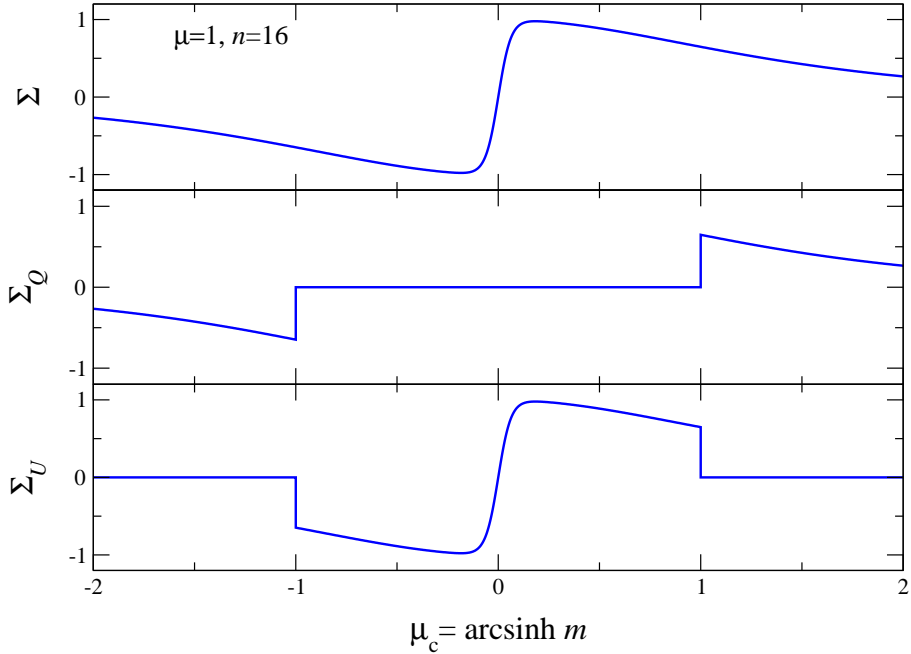} 
\end{center}
 \caption{The full $\mu$-independent chiral condensate (top) in 
one-dimensional QCD along with the contributions from the constant 
(middle) and oscillating part (bottom) of the eigenvalue density. As in 
four-dimensional QCD, at nonzero $\mu$ the discontinuity of the chiral 
condensate at zero quark mass in the thermodynamic limit is entirely due 
to the complex oscillating part of the eigenvalue density (compare to 
Fig.\ 2 of 
Ref.\ \cite{Osborn:2005cy}).
 }
\label{fig:silverblaze}
\end{figure}

In one dimension the essential properties of the eigenvalue density do
not depend on $N_c$. For all $N_c$ the eigenvalues lie on an ellipse in
the complex plane and the eigenvalue density is a rapidly oscillating
complex function with a diverging amplitude in the thermodynamic limit
(when $|\mu|>|\mu_c|$). We continue therefore with the case $N_c=1$,
for which the spectral density has a simple analytical form
\cite{Ravagli:2007rw}. 
If the ellipse is parametrized by an angle $\alpha$, i.e.,
\be
z = \half\left( e^{i\alpha+\mu} - e^{-i\alpha-\mu}\right),
\label{zalr}
\ee
the eigenvalue density becomes \cite{Ravagli:2007rw}, up to an overall 
constant which cancels in Eq.~(\ref{eq:211}) below, 
\be
\rho(\alpha;\mu) = 1-\frac{\cosh[n(\mu+i\alpha)]}{\cosh(n\mu_c)}.
\ee
Observe that the eigenvalue density is complex and is
highly oscillating when $|\mu|>|\mu_c|$, as illustrated in 
Fig.\ \ref{fig:dist}.
 The chiral condensate is then written as
\be
\label{eq:211}
\Sigma = \int_0^{2\pi}\frac{d\alpha}{2\pi}\, 
\rho(\alpha;\mu)\Sigma(\alpha;\mu),
\ee
with
\be
\Sigma(\alpha;\mu) =  \frac{1}{\sinh(\mu+i\alpha)+\sinh(\mu_c)},
\ee
and is evaluated as
\be
\label{eq:Sigmaexact}
\Sigma   = \frac{\tanh(n\mu_c)}{\cosh(\mu_c)},
\ee
in agreement with Eq.\ (\ref{eq:27}) for $N_c=1$.
In the thermodynamic limit the discontinuity when $\mu_c$ goes 
through zero appears and
\be
\lim_{n\to \infty} \Sigma = \frac{\sign(\mu_c)}{\cosh(\mu_c)}.
\ee

As was first observed in the microscopic limit of four-dimensional QCD 
\cite{Osborn:2005ss}, the strong complex oscillations of the eigenvalue 
density are responsible for the discontinuity of the chiral condensate at 
zero quark mass in the thermodynamic limit. To see this it is advantageous 
to split the contribution to the chiral condensate into the contribution 
from the smooth part of the density (the ``1''),
 \be
\Sigma_Q = \int_0^{2\pi}\frac{d\alpha}{2\pi}\,  \Sigma(\alpha;\mu) 
= \Theta\left(|\mu_c|-|\mu|\right) \frac{\sgn(\mu_c)}{\cosh(\mu_c)},
\ee
and the contribution from the complex oscillating part,
\be
\Sigma_U = \int_0^{2\pi}\frac{d\alpha}{2\pi}\, 
\left[\rho(\alpha;\mu)-1\right]\Sigma(\alpha;\mu) = \Sigma - \Sigma_Q.
\ee
 This is illustrated in Fig.\ \ref{fig:silverblaze}. The smooth part 
contributes only for large quark mass (i.e.\ $|\mu_c|>|\mu|$) while the 
oscillating part makes up the chiral condensate when $|\mu_c|<|\mu|$. In 
particular, the oscillations of the eigenvalue density are responsible for 
the discontinuity of the chiral condensate when the quark mass goes 
through zero and provide the solution to the Silver Blaze problem. We will 
demonstrate below that complex Langevin correctly evaluates the 
contribution from these oscillations.

\section{Complex Langevin dynamics}
\label{sec:CL}
\setcounter{equation}{0}

\begin{figure}[t]
\begin{center}
\epsfig{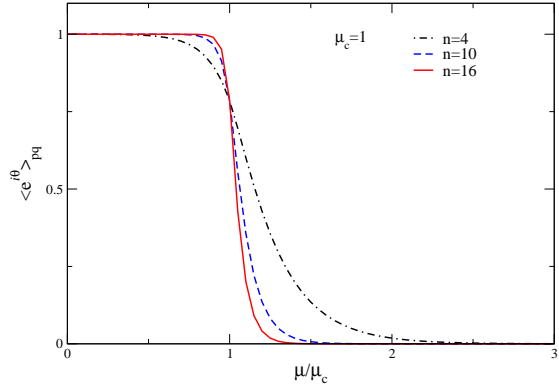}
\end{center}
\caption{
 Expectation value of the phase factor $e^{i\theta} = 
\rho(\alpha;\mu)/|\rho(\alpha;\mu)|$ with respect to the phase quenched 
weight as a function of $\mu/\mu_c$ for $\mu_c=1$ and $n=4,10,16$. 
 }
\label{fig:avphase}
\end{figure}

We interpret $\rho(\alpha;\mu)$ as the complex weight, satisfying the 
usual relation $\rho^*(\alpha;\mu)  = \rho(\alpha;-\mu^*)$, and write it 
as
 \be
\rho(\alpha;\mu) = |\rho(\alpha;\mu)| e^{i\theta}.
\ee
 The severity of the sign problem can be assessed via the expectation 
value of the phase $e^{i\theta}$ with respect to the phase quenched 
weight. We hence define the average phase factor as
 \be 
 \label{eq:phav}
 \bra e^{i\theta}\ket_{\rm pq} =  
 \frac{\int_0^{2\pi}d\alpha\, \rho(\alpha;\mu)}{
 \int_0^{2\pi}d\alpha\, |\rho(\alpha;\mu)|},
\ee
  %%%%%%%%%
  % \be 
  %\bra e^{i\theta}\ket_{\rm pq} =  
  %\frac{\int_0^{2\pi}\frac{d\alpha}{2\pi}\, \rho(\alpha;\mu)}{
  %\int_0^{2\pi}\frac{d\alpha}{2\pi}\, |\rho(\alpha;\mu)|}.	
  %\ee
  % \be 
  %\bra e^{i\theta}\ket_{\rm pq} =  \frac{\Xi_{\rm full}}{\Xi_{\rm pq}},
  %\Xi_{\rm full} = \int_0^{2\pi}\frac{d\alpha}{2\pi}\, \rho(\alpha;\mu),
  %\;\;\;\;\;\;\;
  %\Xi_{\rm pq} = \int_0^{2\pi}\frac{d\alpha}{2\pi}\, |\rho(\alpha;\mu)|.
  %\ee
  %$\Xi_{\rm full}=1$. 
  %%%%%%%%%
  which is shown in Fig.\ \ref{fig:avphase} for various values of $n$. 
 We find the sign problem to be mild (absent) when $|\mu|<|\mu_c|$ and 
severe when $|\mu|>|\mu_c|$. All the dependence on $\mu$ in Fig.\ 
\ref{fig:avphase} emerges from the denominator in Eq.\ (\ref{eq:phav}), 
since the numerator is $\mu$-independent. The theory with the phase 
quenched weight therefore has a transition at $\mu=\mu_c$: our model 
behaves exactly as QCD in the region where $0\leq \mu\lesssim m_B/3$, with 
$\mu_c$ playing the role of $m_\pi/2$.

 The complex weight $\rho(\alpha;\mu)$ is a highly oscillatory complex 
function, with period $2\pi/n$.  
From its real and imaginary parts,
 \bse
\begin{align}
\re\rho(\alpha;\mu) & =  1-\frac{\cosh(n\mu)}{\cosh(n\mu_c)}\cos(n\alpha),
\\
\im\rho(\alpha;\mu) & = \frac{\sinh(n\mu)}{\cosh(n\mu_c)}\sin(n\alpha),
\end{align}
\ese
 we note that the sign problem is severe when the real part of the 
distribution is not positive-definite and, for large $n$, the amplitude of 
oscillations grows exponentially as $\exp[n(|\mu|-|\mu_c|)]$. On the other 
hand, when the sign problem is absent, the amplitude of oscillations 
decreases exponentially in the thermodynamic limit.

We now apply complex Langevin dynamics to study QCD in one dimension. We
interpret $\rho(\alpha;\mu)$ as the distribution and
 \be 
S(\alpha;\mu) = -\log\rho(\alpha;\mu)
\ee 
 as the complex action. We complexify the angle $\alpha\to x+iy$.  
 The discretized Langevin equations, for general complex noise, 
read
\begin{subequations}
\begin{align}
 x_{j+1} &=  x_j +\eps K_x(x_j,y_j) +\sqrt{\eps N_\rmR}\eta_j^\rmR, \\
 y_{j+1} &= y_j +\eps K_y(x_j,y_j) +\sqrt{\eps N_\rmI}\eta_j^\rmI,
 \end{align}
\end{subequations}
where $\eps$ is the Langevin stepsize and Langevin time is 
$\vartheta=j\eps$. 
The drift terms are determined by
\be
 K_x = -\re\frac{\partial S}{\partial \alpha}\Big|_{\alpha\to x+iy}, 
\;\;\;\;\;\;\;
 K_y = -\im\frac{\partial S}{\partial \alpha}\Big|_{\alpha\to x+iy},
\ee
with the classical drift term
\be
 \frac{\partial S}{\partial\alpha}\Big|_{\alpha\to x+iy} 
= in \frac{\sinh[n(\mu-y+ix)]}{\cosh(n\mu_c)-\cosh[n(\mu-y+ix)]}.
\ee
The noise satisfies
\be
\bra \eta_j^{\rmR}\ket = \bra \eta_j^{\rmI}\ket 
= \bra\eta_j^\rmR\eta_{j'}^\rmI\ket =0,
\;\;\;\;
\bra\eta_j^\rmR\eta_{j'}^\rmR\ket = \bra\eta_j^\rmI\eta_{j'}^\rmI\ket 
= 2\delta_{jj'},
\ee
 with $N_\rmR-N_\rmI=1$. Finally, the Fokker-Planck equation underlying 
this stochastic process reads, in the limit that $\eps\to 0$,
 \be
\label{eq:FP}
 \partial_\vartheta P(x,y;\vartheta) = \left[
\partial_x\left(N_\rmR\partial_x
-K_x\right) +\partial_y\left(N_\rmI\partial_y-K_y\right)\right]
P(x,y;\vartheta),
 \ee
 where $P(x,y;\vartheta)$ is a distribution in the complexified space, 
which should be real and positive.
 We specialize to real noise ($N_\rmI=0$) in most of the paper, but 
briefly come back to complex noise at the end.

\begin{figure}[t]
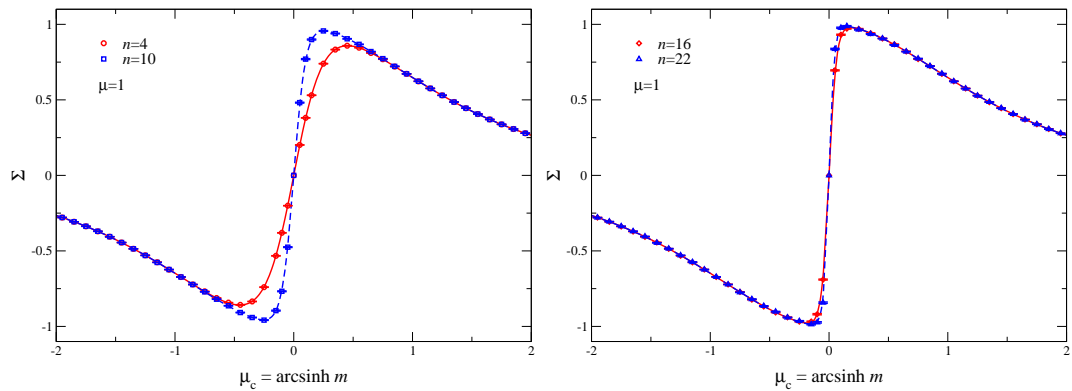

\begin{center}
\epsfig{figure=plot_Sigma_mu1_n4-10.eps,height=5.1cm} 
\epsfig{figure=plot_Sigma_mu1_n16-22.eps,height=5.1cm}
\end{center}
\caption{
 Condensate for $n=4, 10$ (left) and $n=16, 22$ (right) as a function of 
$\mu_c$ for fixed $\mu=1$. The data points are obtained with complex 
Langevin dynamics, while the lines are the exact result 
(\ref{eq:Sigmaexact}).
 }
\label{fig:SigmaCL}
\end{figure}

 We have solved the Langevin equations numerically, using a moderate 
stepsize of $\eps=0.001$ and Langevin times up to $\vartheta=5\times 
10^4$. The 
results for the condensate are shown in Fig.\ \ref{fig:SigmaCL} for four 
values of $n$. We observe excellent agreement with the exact results, 
indicated by the lines. Note that the sign problem is severe when 
$-1<\mu_c<1$. The discontinuity at $\mu_c=0$ emerges in the thermodynamic 
limit. We have done simulations for various values of $\mu$ to confirm 
that the condensate only depends on $\mu_c$ and not on $\mu$.

\section{One stationary distribution}
\label{sec:stat}
\setcounter{equation}{0}

In order to understand why complex Langevin dynamics has no apparent 
problem with the sign and Silver Blaze problems, we first note that $\mu$ 
can be eliminated completely from the dynamics by writing $y=\mu+\bar y$, 
which removes $\mu$ from the classical drift term. The Silver Blaze 
problem is therefore trivially solved by the complexification. Taking this 
one step further, we specialize to the case $y=\mu$. The corresponding 
drift term
 \be
\tilde K \equiv 
 -\frac{\partial S}{\partial\alpha}\Big|_{\alpha\to x+i\mu} 
= n \frac{\sin(nx)}{\cosh(n\mu_c)-\cos(nx)},
\ee
 is entirely real, such that there is no dynamics in the imaginary 
direction (in the case of real noise). We can then interpret the remaining 
Langevin evolution as a real process, shifted in the complex plane, with a 
force $\tilde K$ which can be derived from an action,
 \be
\tilde S = -\log[\cosh(n\mu_c)-\cos(nx)] + \mbox{constant},
\ee
 such that $\tilde K=-\partial \tilde S/\partial x$. It follows that the 
associated probability distribution can be written as\footnote{See also 
Refs.\ \cite{Aarts:2009uq,Lombardo:2009rt,deForcrand:2010ys} for similar 
cases.}
 \be
P_\delta(x,y) = p_x(x)\delta(y-\mu),
\ee
 with
 \be
 p_x(x) = \exp(-\tilde S), \;\;\;\;\;\;\;\; \tilde S(x) =  S(x;\mu=0),
\ee
where the constant has been fixed by the normalization condition
\be
 \int\frac{dxdy}{2\pi}\, P(x,y)=1.
 \ee
 This distribution is related to the original complex weight as
 \be
 p_x(x) = \rho(x;\mu=0),
\ee
and is real and positive.

This distribution gives the correct chiral condensate, since
\bea
 \Sigma = &&\hm  \int_0^{2\pi}\frac{dx}{2\pi}\int_{-\infty}^\infty dy\, 
P_\delta(x,y)\Sigma(x+iy;\mu) 
 \nn \\
 = &&\hm \int_0^{2\pi}\frac{dx}{2\pi} \, p_x(x)\Sigma(x+i\mu;\mu)
 \nn \\ 
= &&\hm \int_0^{2\pi}\frac{dx}{2\pi} \, \rho(x;0)\Sigma(x;0) =  
\frac{\tanh(n\mu_c)}{\cosh(\mu_c)}.
\eea
 The solution we found is indeed a stationary solution of the 
Fokker-Planck equation (\ref{eq:FP}) when $N_\rmI=0$, since 
$K_y\delta(y-\mu) = 0$ and $\tilde K=K_x(\mu=0)$. 
 To conclude, we have found a stationary distribution in the complexified 
space. If a trajectory is initialized at $y=\mu$, we can consider a real 
Langevin process for which standard arguments can be used to demonstrate 
that the correct stationary distribution is reached in the limit of 
infinite Langevin time.

\begin{figure}[t]
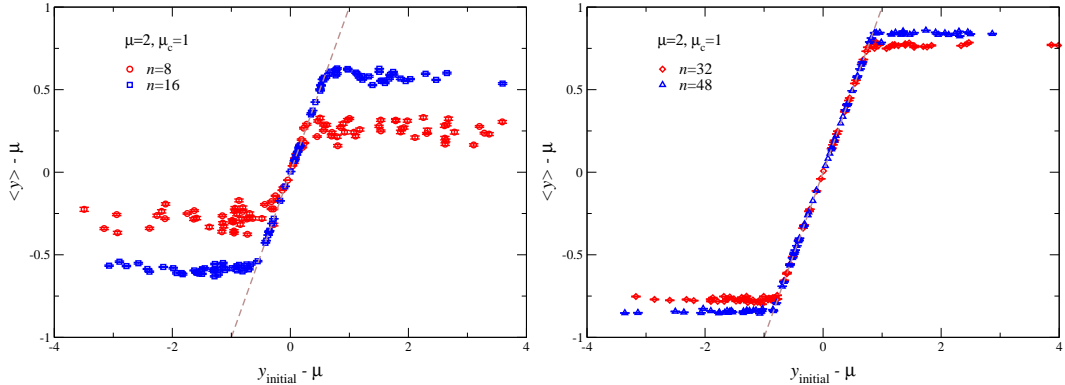

\begin{center}
\epsfig{figure=plot_yyv2.eps,height=5.1cm} 
\epsfig{figure=plot_yyv3.eps,height=5.1cm}
\end{center}
\caption{
 Expectation value $\bra y\ket - \mu$ as a function of (random) initial 
condition 
$y_{\rm initial}-\mu$ for fixed $\mu=2$, $\mu_c=1$ for $n=8,16$ (left) and 
$32, 48$ (right). The dashed lines indicate $\bra y\ket=y_{\rm initial}$.
 }
\label{fig:yav}
\end{figure}

Surprisingly, $P_\delta(x,y)$ is not the unique stationary distribution: 
trajectories initialized with $y\neq \mu$ are not in general attracted to 
$y=\mu$.  To illustrate this, we have computed the expectation value $\bra 
y\ket$ for a large number of trajectories, starting from different initial 
conditions $y_{\rm initial}$. The resulting $\bra y\ket$ is shown in 
Fig.~\ref{fig:yav} as a function of $y_{\rm initial}$ for $n=8, 16$ (left) 
and $n=32, 48$ (right), using 100 initial conditions randomly distributed 
around $y_{\rm initial}=\mu$ for each $n$. Note that we have subtracted 
$\mu=2$ from both $\bra y\ket$ and $y_{\rm initial}$. If all trajectories 
are attracted to the stationary distribution $P_\delta$, one should obtain 
$\bra y\ket=\mu$ independent of $y_{\rm initial}$. Instead we find that 
the average value of $y$ is linearly correlated with the initial value 
when $y_{\rm initial}-\mu$ is small and independent of the initial value 
when $y_{\rm initial}-\mu$ is larger and approaches $\pm \mu_c$. 
Despite this, all 
trajectories yield a value for the condensate that is consistent with the 
exact result.\footnote{Taking the average of the 100 initial conditions we 
find $\Sigma = 0.6479(2)$ for $n=8$, $0.6483(3)$ for $n=16$, $0.6483(4)$ 
for $n=32$ and $0.6478(3)$ for $n=48$. The exact result is 
$\Sigma=0.648054$.} The $y$ value where the crossover between the linear 
dependence on and the independence of the initial condition occurs, 
depends on $n$. From the numerical results we infer that in the 
thermodynamic limit
 \be
\lim_{n\to\infty} \bra y\ket = 
\begin{cases}
 \mu+|\mu_c| & \mbox{when} \;\; y_{\rm initial} > \mu+|\mu_c|, \\
 y_{\rm initial} &  \mbox{when} \;\; \mu-|\mu_c|< y_{\rm initial} < 
\mu+|\mu_c|, \\
 \mu-|\mu_c| & \mbox{when} \;\; y_{\rm initial} < \mu-|\mu_c|.
\end{cases}
\ee
 We conclude that the dynamics is not ergodic. Nevertheless, the 
expectation values of the condensate are consistent within the error with 
the analytical result for all initial conditions. The stationary 
distribution at $y=\mu$ is only realized when $y_{\rm initial}=\mu$.

\section{Classical flow and degenerate distributions}
\label{sec:flow}
\setcounter{equation}{0}

In this section we explain the numerical results observed above. We first 
show analytically why the dynamics is not ergodic and subsequently 
demonstrate that a continuum of distributions exist in the thermodynamic 
limit, all yielding the correct condensate.

\begin{figure}[t]
\begin{center}
\epsfig{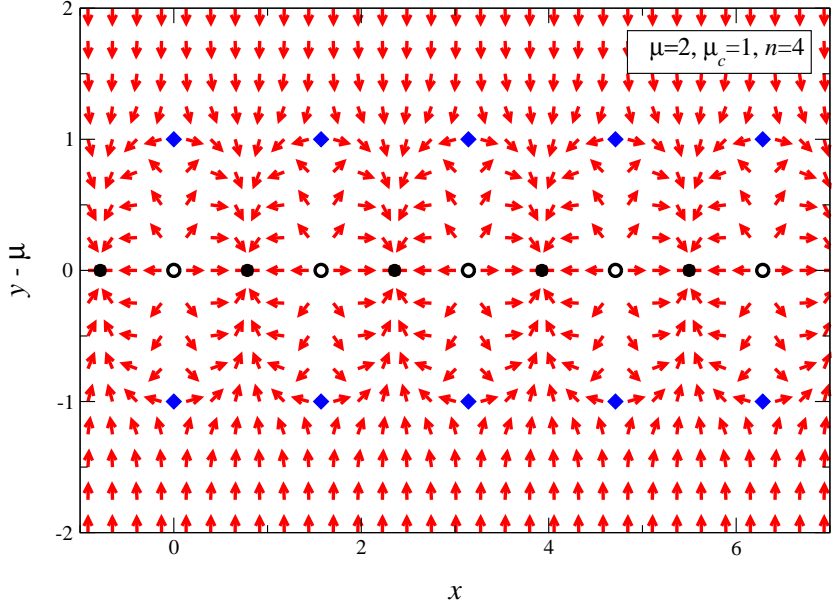}
\end{center}
\caption{
 Flow pattern for $\mu=2$, $\mu_c=1$ and $n=4$.
The filled/open black dots are stable/unstable fixed points, the diamonds 
indicate a diverging flow. On the line $y=\mu$, $K_y=0$.
The arrows are normalized to have the same length.
 }
\label{fig:flow}
\end{figure}

 The nonergodicity can be understood from the classical flow. We  split 
the force explicitly in real and imaginary parts and write 
\be
K_x = n \frac{AE-C}{(E-AC)^2+B^2D^2} D, 
\;\;\;\;\;\;\;\;
K_y = n \frac{A-CE}{(E-AC)^2+B^2D^2} B,
\ee
in terms of
 \bse
\begin{align}
& A = \cosh[n(\mu-y)], & C = \cos(nx),  \\
& B = \sinh[n(\mu-y)], & D = \sin(nx), & &E = \cosh(n\mu_c).
\end{align}
\ese
Classical fixed points are determined by $K_x = K_y = 0$.
We find
\begin{itemize}
\item[$-$] $n$ stable fixed points at
\be
 x=(2k+1)\pi/n, \;\; y=\mu \;\;\;\; (k=0,\ldots,n-1).
 \ee
 \item[$-$] $n$ unstable fixed points at
\be
 x=2k\pi/n, \;\; y=\mu \;\;\;\; (k=0,\ldots,n-1).
 \ee
 \item[$-$] $2n$ points where the flow diverges ($K_x = 0, K_y = \infty$) 
at
\be
 \label{eq:div}
 x=2k\pi/n, \;\; y=\mu\pm\mu_c \;\;\;\; (k=0,\ldots,n-1).
 \ee
 \end{itemize}
 The flow patterns are shown in Fig.\ \ref{fig:flow} for $n=4$. For 
larger values of $n$, the number of fixed points and hence 
the density of regions where the flow changes direction increase. The 
region bounded by $\mu\pm\mu_c$ is an attractor region: all trajectories 
will end up here, irrespective of the initial $y$ value. The line $y=\mu$, 
however, is not an attractor due to the alternating stable and unstable 
fixed points, except when starting exactly on it. For obvious reasons, we 
will refer to the region bounded by $y=\mu\pm\mu_c$ as the inside region, 
while the two regions where $|y-\mu|>|\mu_c|$ are referred to as the 
outside regions.

\begin{figure}[t]
\begin{center}
\epsfig{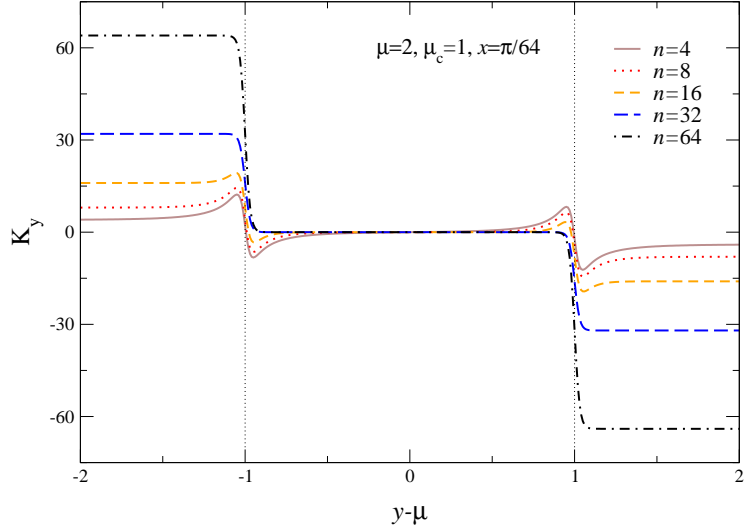} 
\end{center}
\caption{
 Force $K_y$ in the $y$-direction as a function of $y-\mu$ at fixed 
$x=\pi/64$ for various values of $n$ and $\mu=2$, $\mu_c=1$.
 }
\label{fig:Ky}
\end{figure}

The arrows in Fig.\ \ref{fig:flow} are normalized to have the same length. 
To indicate the strength of the force in $y$ direction, we show in Fig.\ 
\ref{fig:Ky} the value of $K_y$ at fixed $x=\pi/64$. We observe that in 
the thermodynamic limit $K_y$ goes to zero in the inside region, 
whereas its magnitude increases linearly with $n$ in the outside region.
 This is confirmed by the following expressions for $K_y$ in the 
thermodynamic limit,\footnote{\label{ft1} Exactly on the borderlines 
$y=\mu\pm\mu_c$, 
$K_y$ diverges for specific values of $x$, see Eq.\ (\ref{eq:div}). Away 
from these points, the force is directed inwards.}
 \be
\lim_{n\to\infty} K_y  = 
\begin{cases} 
n\,\sign(y-\mu) e^{-n(|\mu_c|-|\mu-y|)} \cos(nx) \to 0, &\;\; 
\mbox{inside,}  \\
n\,\sign(\mu-y) \to \pm\infty, & \;\;  \mbox{outside}. 
\end{cases}
\ee
On the other hand, the force in the $x$ direction goes to zero both 
on the inside and the outside, 
\be
\lim_{n\to\infty} K_x  = 
\begin{cases} 
 n e^{-n(|\mu_c|-|\mu-y|)}\sin(nx) \to 0, & \;\; \mbox{inside}, \\
 n e^{-n(|\mu-y|-|\mu_c|)} \sin(nx) \to 0, & \;\; \mbox{outside}. 
\end{cases}
\ee
 This confirms that for large $|y|$ the flow is attracted to the region 
$\mu-|\mu_c|<y<\mu+|\mu_c|$ very efficiently. Once inside, the forces 
vanish exponentially, with the rate determined by the vicinity to the 
boundary at $\pm\mu_c$. This explains the dependence on initial conditions 
found in Fig.\ \ref{fig:yav}.

In the thermodynamic limit, the forces in the inside region vanish. The 
stochastic evolution then reduces to simple diffusion in a square well 
bounded by $y=\mu\pm\mu_c$ (and periodic in $x$). In the case of real 
noise the diffusion is one-dimensional. In the case of complex noise, the 
diffusion is two-dimensional. 
 It is now straightforward to deduce the stationary solution of the 
Fokker-Planck equation in the thermodynamic limit. We find
\be
\lim_{n\to\infty} P(x,y) =
\begin{cases}
p_y(y) & \mbox{when} \;\; \mu-|\mu_c|< y < \mu+|\mu_c|, \\
0    & \mbox{elsewhere}.
\end{cases}
\ee
In the case of real noise, $p_y(y)$ depends on the initial conditions 
as
 \be
 p_y(y) =
 \begin{cases} 
 \delta(y-\mu-|\mu_c|-\vareps)
 & \mbox{when} \;\; y_{\rm initial} \geq \mu+|\mu_c|, \\
 \delta(y-y_{\rm initial}) & \mbox{when} \;\; \mu-|\mu_c|< y_{\rm initial} 
< \mu+|\mu_c|, \\
 \delta(y-\mu+|\mu_c|+\vareps) & \mbox{when} \;\; y_{\rm initial} \leq 
\mu-|\mu_c|,
 \end{cases}
\ee
where $\vareps\downarrow 0$, see footnote \ref{ft1}.
 For complex noise, $p_y(y)$ is determined by the normalization condition 
as
\be
p_y(y) = \frac{1}{2|\mu_c|}.
\ee
 The resulting distribution is sketched in Fig.\ \ref{fig:dist}.

\begin{figure}[t]
\begin{center}
\epsfig{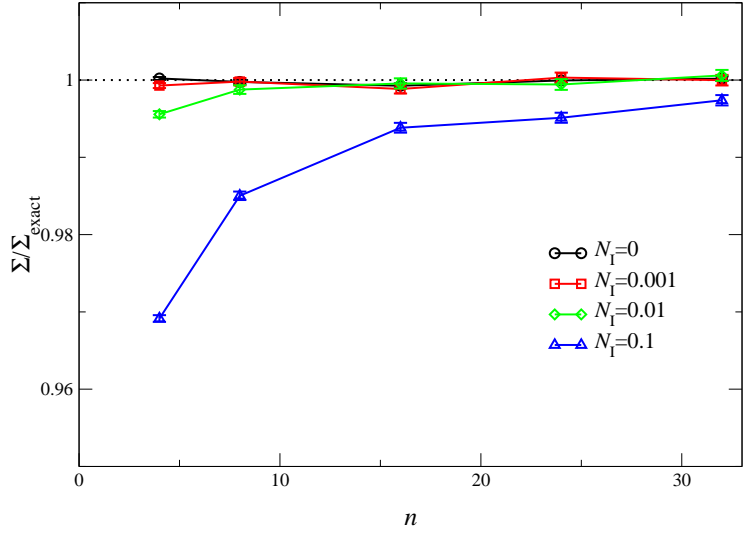}
\end{center}
\caption{
 Normalized condensate $\Sigma/\Sigma_{\rm exact}$ as a function of $n$ in 
the case of complex noise, $N_\rmI>0$. It is seen that in this case 
complex noise works in the thermodynamic limit, $n\to \infty$.
 }
\label{fig:NI}
\end{figure}

This continuous family of distributions all yields the correct value for 
the condensate. To demonstrate this, we write the condensate as
\be
\label{eq:511}
\Sigma = \int\frac{dxdy}{2\pi}\, P(x,y)\Sigma(x+iy;\mu) =
\int dy \, p_y(y) \int\frac{dx}{2\pi}\, \Sigma(x+iy;\mu),
\ee
and evaluate the $x$ integral by contour integration. In 
terms of $z=e^{ix}$, we find
\be
\int\frac{dx}{2\pi}\, \Sigma(x+iy;\mu) = \int_{|z|=1}\frac{dz}{2\pi i}
\frac{-2e^{-\mu+y}}{(z-z_1)(z-z_2)},
\ee
with
\be
z_1 = e^{\mu_c-\mu+y},
\;\;\;\;\;\;\;\;\;
z_2 = - e^{-\mu_c-\mu+y}.
\ee
In the inside region, $\mu-|\mu_c|< y < \mu+|\mu_c|$, only the pole at 
$z_2$ ($z_1$) contributes when $\mu_c>0$ ($\mu_c<0$). The result is
\be
\int\frac{dx}{2\pi}\, \Sigma(x+iy;\mu)  = 
\frac{\sgn(\mu_c)}{\cosh(\mu_c)},
\ee
 for all values of $y$ in the inside region. The remaining integral over 
$y$ in Eq.\ (\ref{eq:511}) is now trivially performed and yields unity due 
the normalization 
condition. We conclude therefore that the correct result for the 
condensate is obtained and that the degenerate distributions are all 
equivalent.

It is known that complex noise does not work in general 
\cite{Aarts:2009uq} and that is also what we find here for finite $n$. 
However, in this example complex noise can be expected to work in the 
thermodynamic limit, since in that case the dynamics takes place in a 
square well with infinitely-high walls at $y=\mu\pm\mu_c$. This is 
demonstrated in Fig.~\ref{fig:NI}, where the condensate is shown as a 
function of $n$ for various values of $N_\rmI$.

\section{Conclusion}
\label{sec:concl}
\setcounter{equation}{0}

We have established a first link between the complex oscillations of the 
universal microscopic spectral density of the Dirac operator and complex 
Langevin dynamics. For QCD in one dimension we have shown how complex 
Langevin dynamics correctly evaluates the chiral condensate given the 
complex and strongly oscillating unquenched eigenvalue density. The exact 
solution of the Fokker-Planck equation in the thermodynamic limit shows 
explicitly how the complex Langevin method can deal with the severe sign 
problem present. Surprisingly we did not find a unique solution but rather 
a continuum of degenerate solutions, which all yield the correct chiral 
condensate. This has been shown analytically in the thermodynamic limit 
and demonstrated numerically for finite systems.

The exact solution presented here offers a direct analytic indication that 
complex Langevin dynamics can solve the sign problem. While there exist 
examples where the complex Langevin method is problematic 
\cite{Ambjorn:1986fz,Aarts:2009uq,Aarts:2010aq}, we would like to stress 
that the difficulties encountered with complex Langevin dynamics are 
independent of the severity of the sign problem. For instance, in Ref.\ 
\cite{Aarts:2010aq} it was demonstrated that the failure of complex 
Langevin dynamics is caused by an 
apparent incorrect exploration of the complexified field space by the 
Langevin evolution, similar to the case of complex noise considered in 
Ref.\ \cite{Aarts:2009uq}. In the cases where the method works well, such 
as in Refs.\ \cite{Aarts:2008wh,Aarts:2009hn} and above, the thermodynamic 
limit poses no obstacle. This is in strong contrast to the standard 
methods (reweighting, Taylor series, imaginary chemical potential and 
analytic continuation), which work well in small volumes but eventually 
break down due the sign problem in large volumes. Our findings therefore 
strongly encourage further studies of complex Langevin dynamics at nonzero 
chemical potential.

\vspace*{0.5cm}
\noindent
{\bf Acknowledgments} 
Part of this work was carried out at the Yukawa 
Institute for Theoretical Physics in Kyoto. It is a pleasure to thank the 
Yukawa Institute, and especially Kenji Fukushima, for hospitality.  
We thank 
Poul Henrik Damgaard, 
Philippe de Forcrand, 
Simon Hands, 
Frank James, 
Erhard Seiler,
Ion-Olimpiu Stamatescu 
and
Jac Verbaarschot
for discussions. 
Finally, we are grateful to the Niels Bohr Institute for its 
hospitality during the completion of this work. 
The work of G.A.\ is supported by STFC. 
The work of K.S.\ is funded by the Danish Natural Science Research 
Council.

\end{document}